\documentclass[twocolumn,letter]{IEEEtran}
\usepackage[usenames,dvipsnames]{color}
\usepackage{subfigure}
\usepackage{graphicx}
\usepackage{amsmath}
\usepackage{color}
\usepackage{multicol}
\usepackage{amssymb}

\usepackage{amsfonts}%
\usepackage{verbatim}%
\usepackage{enumerate}%
\usepackage{psfrag}
\usepackage{epsfig}
\usepackage{multicol}
\usepackage{setspace}
\usepackage{dsfont}

\usepackage{balance}

\usepackage{algorithm}
\usepackage{algorithmic}
\usepackage{cite}
\usepackage{float}
\usepackage{cite}
        \newcommand\blfootnote[1]{%
  \begingroup
  \renewcommand\thefootnote{}\footnote{#1}%
  \addtocounter{footnote}{-1}%
  \endgroup}
 
\begin{document}
\title{Protocol Design and Stability Analysis of Cooperative Cognitive Radio Users}
\author{ Ahmed El Shafie$^\dagger$, Tamer Khattab$^*$ and H. Vincent Poor$^{**}$\\
\small \begin{tabular}{c}
$^\dagger$Wireless Intelligent Networks Center (WINC), Nile University, Giza, Egypt. \\
$^*$Electrical Engineering, Qatar University, Doha, Qatar. \\
$^{**}$ Electrical Engineering, Princeton University, Princeton, NJ, USA.
\end{tabular}
}
\date{}
\maketitle
\thispagestyle{empty}
\pagestyle{empty}
\blfootnote{This paper was made possible by a NPRP grant 09-1168-2-455 from the
Qatar National Research Fund (a member of The Qatar Foundation). The
statements made herein are solely the responsibility of the authors.}
\begin{abstract}
A single cognitive radio transmitter--receiver
pair shares the spectrum with two primary users communicating with their respective receivers.
Each primary user has a local traffic queue, whereas the cognitive user has three queues; one storing its own traffic while the other two are relaying queues used to store primary relayed packets admitted from the two primary users. A new cooperative cognitive medium access control protocol for the described network is proposed, where the cognitive user exploits the idle periods of the primary spectrum bands. Traffic arrival to each relaying queue is controlled using a tuneable admittance factor, while relaying queues service scheduling is controlled via channel access probabilities assigned to each queue based on the band of operation. The stability region of the proposed protocol is characterized shedding light on its maximum expected throughput. Numerical results demonstrate the performance gains of the proposed cooperative cognitive protocol.
\end{abstract}

\section{Introduction}
The quest for efficient radio spectrum utilization and high performance wireless communication networks relies, among other technologies, on cooperative communications and cognitive radios. Cognitive relaying, which involves cooperation among primary and secondary terminals, has been investigated in \cite{simeone,khattab,krikidis2009protocol,close,krikidis2010stability,bao2010stable,Sult1212:Cooperative}. 

In \cite{simeone}, a cognitive radio transceiver is used as a relay for the undelivered packets of the primary user. The secondary user aims at maximizing its mean service rate subject to the stability of all other queues in the network via optimizing its own power.
In \cite{khattab}, the authors consider a cognitive setting with one primary user and one secondary user. The secondary user aims at minimizing its average delay by tuning the relayed fraction of the primary undelivered packets. The authors of \cite{krikidis2009protocol} propose a cluster of secondary users helping the primary transmitter with a single relaying queue accessible by all the secondary users. In \cite{close}, the authors investigate an extension of the cooperation problem with multiple secondary transmitters acting as relays for the undelivered packets of the primary transmitter. Furthermore, the authors of \cite{close} consider priority in transmission given to the relaying queues of the cognitive radio users. In \cite{krikidis2010stability}, Krikidis {\it et al.} consider a network in which one cognitive radio user shares the spectrum with two primary users
wishing to communicate to a single receiver in a multi-access
channel (MAC). The secondary terminal relays the undelivered packets of the primary users stored in the relaying queues. The authors consider a priority of transmission given to the relaying queues over the secondary user's own queue when the primary queues are empty. Furthermore, the cognitive radio user transmits its own packets in two ways: 1) when all the primary and relaying queues are empty; or 2) simultaneously with the primary users via a superposition technique when the primary queues are nonempty. In \cite{bao2010stable}, the authors consider multiple primary users with a common destination and one cognitive radio user with relaying capability. The cognitive radio user sends packets from relaying queues until all are emptied. Afterwards, the cognitive radio user switches to the idle band with the highest instantaneous channel gain to send its own packets. In \cite{Sult1212:Cooperative}, El Shafie {\it et al.} propose a novel multiple access protocol in which the cognitive nodes are ordered in terms of channel access. The cognitive nodes are capable of relaying the undelivered primary packets.

In this work, we investigate a cognitive radio scenario with one cognitive user and two primary users. Unlike most of the existing works, we assume that the primary users operate using frequency-division multiple-access (FDMA). In contrast to \cite{krikidis2010stability}, we do not assume higher priority assigned to relaying packets over the secondary own queue when the primary queues are empty. Also, we relax the assumptions of symmetric primary users and a common receiver. We do not assume the availability of channel state information (CSI) at transmitters. Moreover, we consider different packet sizes for each node in the system. In contrast with previous work (e.g.~\cite{krikidis2010stability,bao2010stable} and the references therein), a new cognitive protocol is proposed. Under this protocol, the cognitive radio user admits tuneable fractions of the undelivered primary users' packets and assigns access probabilities to each of its relaying queues. These access probabilities are assigned based on the joint state of the primary queues. When primary bands are sensed idle (free of primary users' activity), the cognitive user may exploit one or both bands for the transmission of its own data; or for retransmission of primary relayed packets stored in the relaying queues. This protocol allows the cognitive user to access the channel whenever at least one of the primary users is inactive (at least one band is empty).

It must be noted that the proposed cognitive cooperation protocol and the theoretical development in this work can be readily generalized to networks with more than two primary users and more than one cognitive user, where several primary users may choose one or more cognitive radio users or the best cognitive user for cooperation.

%
This paper is structured as follows: Next we describe the system model adopted. We provide the analysis of the stability region and the problem formulation in Section \ref{secx}. In Section \ref{sec3}, we provide some numerical results. The conclusions are drawn in Section \ref{sec4}.

\section{System Model}
We consider a cognitive scenario with one secondary user (cognitive radio user) and two primary users. The primary users are multiplexing over orthogonal frequency channels using FDMA. Each of the two primary users is assigned a unique orthogonal band. The ${\rm m}${\it th} primary user, ${\rm p_m}$, uses band number ${\rm m}$. Let ${\rm s}$, ${\rm pd_m}$ and ${\rm sd}$ denote the secondary user, the primary destination of user ${\rm p_m}$, and the secondary destination, respectively. The channel and time are slotted and the length of one time slot is $T$ seconds. Each user has an infinite length buffer (queue) to store its own incoming fixed-length packet arrivals, denoted by $Q_{\rm i}$ (see~\cite{krikidis2010stability,close} for a similar assumption). The cognitive user has two additional relaying queues, which store primary users' packets admitted for relaying. Let $Q_{\rm sr_{p_1}}$ and $Q_{\rm sr_{p_2}}$ denote the relaying queues for the undelivered admitted packets from users ${\rm p_1}$ and ${\rm p_2}$, respectively. The
arrivals at queue $Q_{\rm i}$ are assumed to be independent and identically
distributed (i.i.d.) Bernoulli random variables~\cite{krikidis2010stability,close} with mean $\lambda_{\rm i}\in[0,1]$ packets of size $b_{\rm i}$ bits per time slot, where ${\rm i}\in\{{\rm p_1}, {\rm p_2}, {\rm s}\}$ for primary user assigned to band $1$, primary user assigned to band $2$ and cognitive user, respectively. The arrivals are also mutually independent from terminal to terminal.


All wireless links exhibit fading and are corrupted by additive white Gaussian
noise (AWGN). The fading is assumed to be stationary, with
frequency non-selective Rayleigh block fading. This means
that the fading coefficient $h_{\rm i,\ell}$ (channel gain for the link connecting node ${\rm i}\in\{\rm p_1,p_2,s\}$ and node ${\rm \ell}\in\{\rm pd_1,pd_2,s,sd\}$) remains
{\it constant during one time slot and over both primary bands}, but changes independently from one
slot to another according to a circularly symmetric complex
Gaussian distribution with zero mean and variance $\sigma^2_{\rm i,\ell}$. Furthermore, the AWGN at node $\ell$ is assumed to be of
zero mean with variance $\mathcal{N}_\ell$.
The primary node ${\rm p_m}$ has bandwidth ${\rm W_{p_m}}$. We do not assume the availability of CSI at the transmitters.

The data packet of a primary node ${\rm p_m}$ contains $b_{\rm p_m}$ bits. The primary users transmit at the beginning of the time slot for $T$ seconds (slot duration) and each of them transmits over its own bandwidth. Hence, the spectral efficiency of primary user ${\rm p_m}$ is $R_{\rm p_m,pd_m}\!=\!b_{\rm p_m}/(TW_{\rm p_m})$ bits/sec/Hz. The cognitive radio user perfectly senses both bands simultaneously within $\tau$ seconds relative to the beginning of the time slot. Note that the sensing duration, $\tau$, needs to be long enough to justify the perfect sensing assumption.\footnote{A similar assumption of perfect sensing is found in \cite{khattab} and \cite{krikidis2010stability}.} Therefore, we assume that $\tau$ is a non--negligible fraction of the time slot. Since the cognitive radio user spends $\tau$ seconds in spectrum sensing, the remaining time for data transmission is $T-\tau$. The packets of the relaying queue $Q_{\rm sr_{p_m}}$ are of size $b_{\rm p_m}$ bits and have spectral efficiency $R_{\rm s,pd_m}\!=\!b_{\rm p_m}/((T-\tau)W_{\rm p_m})$ bits/sec/Hz, whereas the cognitive user's own data packets are of size $b_{\rm s}$ bits and have spectral efficiency $R_{{\rm s,sd}}\!=\!b_{\rm s}/((T-\tau)W_{\rm s})$ bits/sec/Hz. As will be detailed later, the cognitive radio user can either exploit one of the two primary bandwidths individually ${\rm W_s\!=\!W_{\rm p_m}}$ or {\bf combine} both into an {\it aggregate} band ${\rm W_s\!=\!W\!=\sum_{\rm m=1}^2 W_{\rm p_m}\!=\!W_{\rm p_1}\!+\!W_{\rm p_2}}$. Outage of an arbitrary link connecting any two nodes in the network occurs
when the instantaneous capacity of the link is lower than the transmitted spectral efficiency rate. The packet correct reception is characterized by
the success probability \cite{krikidis2010stability,sadek}
\begin{equation}
\overline{P}_{\rm i\rightarrow \ell, W_{i}}\!=\!{\rm Pr}\bigg\{\!\log_2\big(1 \!+\! \gamma_{\rm i,\ell}|h_{\rm i,\ell}|^2\!\big)\!>\!{R_{{\rm i},\ell}}\bigg\}\!=\!\exp\bigg(\!-\frac{2^{R_{{\rm i},\ell}}\!-\!1}{\sigma^2_{\rm i,\ell}\gamma_{\rm i,\ell}}\bigg),
\label{choutage}
\end{equation}
 where $\overline{\mathcal{X}}\!=\!1\!-\!\mathcal{X}$, $\gamma_{\rm i,\ell}\!=\!\mathbb{P}_{\rm i}/\mathcal{N}_\ell$ is the received average signal-to-noise ratio (SNR), when the channel gain is unity, and $\mathbb{P}_{\rm i}$ is the transmit power of node ${\rm i}$.

A feedback acknowledgement/negative-acknowledgement (ACK/NACK) message is sent from a receiver at the end of each time slot to inform its respective transmitter about the status of its packet decoding. The feedback message is overheard by all nodes in the network. Errors in the feedback messages are negligible, which is reasonable for short length packets as strong and low rate
codes can be employed in the feedback channel \cite{krikidis2010stability,sadek}.
A correctly received packet is removed from the respective transmitter's queue.

\section{Stability Analysis and Problem Formulation}\label{secx}
A fundamental performance measure of a communication network is the stability of the queues. Stability can be defined rigorously as follows: \emph{For every queue represented by an irreducible and aperiodic Markov chain with countable number of states, the chain and its associated queue are called stable if and only if
there is a positive probability for the queue, represented by the chain, to become empty}. Denote by $Q^{\left(t\right)}$ the length of queue $Q$ at the beginning of time slot $t$. $Q$ is said to be stable if $\lim_{x \rightarrow \infty  } \lim_{t \rightarrow \infty  } {\rm Pr}\{Q^{\left(t\right)}<x\}=1$~\cite{sadek}. In a multiqueue system, the system is stable when \emph{all} queues are stable. We can apply the following theorem to check the stability of a queue \cite{sadek}. Loynes' theorem: \emph{if the arrival process and the service process of a queue are strictly stationary, and the average service rate is greater than the average arrival rate of the queue, then the queue is stable. If the average service rate is lower than the average arrival rate, then the queue is unstable}.

The probability of a network queue $Q_{\rm i}$ to be empty is $\pi_{\rm i}\!=\!{\rm Pr}\{Q_{\rm i}\!=\!0\}=1\!-\!\frac{\lambda_{\rm i}}{\mu_{\rm i}}$. Since the primary users are operating under an FDMA scheme with orthogonal bands, the states of the queues are independent, and the joint probability is given by ${\rm Pr}\{Q_{\rm p_1}\!=\!\mathcal{K}_1, Q_{\rm p_2}\!=\!\mathcal{K}_2\}\!=\!{\rm Pr}\{Q_{\rm p_1}\!=\!\mathcal{K}_1\} {\rm Pr}\{Q_{\rm p_2}\!=\!\mathcal{K}_2\}$, where $\mathcal{K}_1$ and $\mathcal{K}_2$ are two positive integers.

 The cognitive radio user chooses to accept an undelivered packet from user ${\rm p_1}$ with probability $\alpha_{\rm sr_1}$ and from user ${\rm p_2}$ with probability $\alpha_{\rm sr_2}$. These probabilities are totally independent of all other events. The cognitive radio user accesses the channel on the basis of the primary queues state $({\rm I}_{Q_{\rm p_1}},{\rm I}_{Q_{\rm p_2}})$, where ${\rm I}_{Q_{\rm p_m}}=1$ if ${\rm p_m}$ is active, and zero if it is inactive. The activity of the primary users is captured by the secondary user via channel sensing.

We assume that the cognitive radio user can transmit and receive at the same time (full duplex) using two distinct frequency bands. We also assume that the cognitive radio user cannot send more than one packet in any time--slot from any of its queues. The cognitive radio user admits a certain fraction of the undelivered packets of each primary user.

\textbf{Intuition:} It may seem that the cooperation process based on the secondary user relaying primary packets only causes the secondary user to waste time--slots for relaying primary packets that could be otherwise used for its own packets. However, it turns out that the secondary user is indeed gaining since opportunistic relaying of primary packets results in emptying
primary queues faster; in return, more network resources can
be utilized for delivering the secondary user's packets. As a result, all
users simultaneously achieve performance gains.

\subsection{Primary User Operation} \label{sec:priop}

When a primary packet is not successfully delivered
to a primary destination, but decoded by the cognitive (secondary) node, it is added
to the relevant relaying queue, and a feedback is broadcast by
the cognitive user acknowledging the reception of the packet.
Accordingly, this packet is removed from the respective primary queue.

A packet at the head of the primary queue, $Q_{\rm p_m}$, is served in either of the following mutually exclusive events: if the link connecting user ${\rm p_m}$ with its respective receiver is not in outage; or if the link is in outage, but the link between user ${\rm p_m}$ and the cognitive radio user is not in outage, and the cognitive radio user decides to accept the packet. The mean service rates of the primary queues are
\begin{equation}
\begin{split}
\mu_{\rm p_m}&=\overline{P}_{\rm  p_m\rightarrow pd_m,W_{\rm p_m}}+P_{\rm  p_m\rightarrow pd_m,W_{\rm p_m}} \overline{P}_{\rm  p_m \rightarrow s,W_{\rm p_m}}\alpha_{\rm sr_m},
\end{split}
\end{equation}

The term $P_{\rm  p_m\rightarrow pd_m,W_{\rm p_m}} \overline{P}_{\rm  p_m \rightarrow s,W_{\rm p_m}}\alpha_{\rm sr_m}\ge 0$ is an additional primary throughput due to the availability of the secondary user to assist, i.e., to relay the primary packets. This term is linear in terms of the acceptance fraction, $\alpha_{\rm sr_m}$, which controls the fraction of accepted primary packet for relaying. One can conclude that cooperation never harms the primary users. Specifically, cooperation will always guarantee non-negative gains to the primary service rates relative to the non-cooperative case.

\subsection{Secondary User Operation} \label{sec:secop}
The operation of the secondary user can be described as follows. At each time--slot, the cognitive radio user senses both bands simultaneously for $\tau$ seconds relative to the beginning of the time--slot. When both primary users' queues are empty, the cognitive radio user may: send a packet from $Q_{\rm s}$ using all the available channel bandwidth, ${\rm W}$, with probability $\eta_1$; send a packet from $Q_{\rm s}$ using bandwidth ${\rm W}_{\rm p_2}$ and another packet from the relaying queue $Q_{\rm p_1}$ using bandwidth ${\rm W}_{\rm p_1}$ with probability $\eta_2$; send a packet from $Q_{\rm s}$ using bandwidth ${\rm W}_{\rm p_1}$ and the other packet from $Q_{\rm p_2}$ using bandwidth ${\rm W}_{\rm p_2}$ with probability $\eta_3$; or send a packet from each of the relaying queues, $Q_{\rm p_1}$ and $Q_{\rm p_2}$, with the relevant bandwidth to the respective receivers, which occurs with probability $\eta_4$. The probabilities $\eta_n$ are probabilities of exhaustive and mutually exclusive events, where $n \in \{1,2,3,4\}$. Therefore, $\sum_{n=1}^4 \eta_n=1$.

A packet from $Q_{\rm s}$ is served if either one of the following events takes place: both primary users are inactive including the aforementioned four scenarios with probabilities $\eta_n$ and the link between cognitive user and its destination, ${\rm sd}$, is not in outage; or only one of the primary users is inactive, in which case the cognitive radio user decides to send a packet from $Q_{\rm s}$ with probability $a_{\rm s_m}$ if queue $Q_{\rm p_m}$ is empty, and the link to its respective destination, ${\rm pd_m}$, is not in outage. The mean service rate of $Q_{\rm s}$ is then given by
\begin{equation}
\begin{split}
\mu_{\rm s}&=\pi_{\rm p_1}\pi_{\rm p_2} \biggr[\eta_1  \overline{P}_{\rm  s \rightarrow sd,W} \!+\!  \eta_2 \overline{P}_{\rm  s \rightarrow sd,W_{\rm p_1}} \!+\! \eta_3  \overline{P}_{\rm  s \rightarrow sd,W_{\rm p_2}}\biggr] \\& \ \!+\! \pi_{\rm p_1} \overline{\pi_{\rm p_2}} a_{\rm s_1}  \overline{P}_{\rm  s \rightarrow sd,W_{\rm p_1}} \! +\!  \overline{\pi_{\rm p_1}}\pi_{\rm p_2} a_{\rm s_2} \overline{P}_{\rm  s \!\rightarrow \!sd,W_{\rm p_2}},
\label{fg1}
\end{split}
\end{equation}
\noindent where $\overline{\pi_{\rm p_m}}\!=\lambda_{\rm p_m}/\mu_{\rm p_m}$ is the probability of the primary queue $Q_{\rm p_m}$ being nonempty, which is the complement of $\pi_{\rm p_m}$.

A packet from the relaying queue $Q_{\rm sr_{p_1}}$ is served if the relevant primary user, ${\rm p_1}$, is inactive, the cognitive radio user chooses to send a packet from $Q_{\rm sr_{p_1}}$, which occurs with probability $a_{\rm sr_1}$ if ${\rm p_2}$ is active and with probability $(\eta_2+\eta_4)$ if ${\rm p_2}$ is inactive, and if the link from the cognitive radio user to the respective destination (i.e., the receiver of user ${\rm p_1}$) is not in outage. Hence, the mean service rate of $Q_{\rm sr_{p_1}}$ is given by
\begin{equation}
\begin{split}
\mu_{\rm sr_1}&\!=\! \pi_{\rm p_1} [\overline{\pi_{\rm p_2}} \ a_{\rm sr_1}+\pi_{\rm p_2} \ (\eta_2+\eta_4)]  \overline{P}_{\rm  s \rightarrow pd_1,W_{\rm p_1}}.
\label{fg2}
\end{split}
\end{equation}
Similarly, the mean service rate of $Q_{\rm sr_{p_2}}$ is given by
\begin{equation}
\begin{split}
\mu_{\rm sr_2}&\!=\! \pi_{\rm p_2} [\overline{\pi_{\rm p_1}} \ a_{\rm sr_2}+\pi_{\rm p_1} \ (\eta_3+\eta_4)]  \overline{P}_{\rm  s \rightarrow pd_2,W_{\rm p_2}}.
\label{fg3}
\end{split}
\end{equation}

From Eqns. (\ref{fg1}), (\ref{fg2}) and (\ref{fg3}), the access probabilities assigned to the secondary queues control their service processes. The secondary user adjusts the access probabilities to achieve the optimal throughput.

We finally investigate the arrival process and the mean arrival rate of the relaying queue $Q_{\rm sr_m}$. A packet arrives at the relaying queue when the primary user ${\rm p_m}$ is active, the channel between ${\rm p_m}$ and ${\rm pd_m}$ is in outage, the channel between ${\rm p_m}$ and the cognitive radio user is not in outage, and the cognitive radio user decides to accept the packet. The mean arrival rate is then given by
\begin{equation}
\begin{split}
  \lambda_{\rm sr_m}\!=\!P_{\rm  p_m\rightarrow pd_m,W_{\rm p_m}}  \overline{P}_{\rm  p_m \rightarrow s,W_{\rm p_m}} \alpha_{\rm sr_m} \ \overline{\pi_{\rm p_m}}.
\end{split}
\end{equation}
We note that $\alpha_{\rm sr_m}$ controls the flow of the undelivered packets of the ${\rm m}${\it th} primary user to the secondary relaying queue $Q_{\rm sr_{p_m}}$. Varying $\alpha_{\rm sr_m}$ from $0$ to $1$ manages the arrival rate of primary packets at $Q_{\rm sr_{p_m}}$.
 The secondary user adjusts $\alpha_{\rm sr_m}$ to achieve the optimal acceptable fraction of the undelivered primary packets from each primary user such that the relaying queues are maintained stable.

The stability region of the proposed system is characterized by the closure of the rate tuples $(\lambda_{\rm s},\lambda_{\rm p_1},\lambda_{\rm p_3})$. An efficient technique to obtain this closure is to solve a constrained optimization problem. The optimization problem aims at obtaining the maximum mean arrival rate to $Q_{\rm s}$ for each feasible $\lambda_{\rm p_1}$ and $\lambda_{\rm p_2}$ as $\alpha_{\rm sr_1},\alpha_{\rm sr_2},a_{\rm s},a_{\rm sr_1}$, $a_{\rm sr_2}$, and $\eta_n \;\;\forall n\in\{1,2,3,4\}$ vary over the set $[0,1]$ with $a_{\rm s_1}+a_{\rm sr_1}=1,\ a_{\rm s_2}+a_{\rm sr_2}=1$ and $\sum_{n=1}^4\eta_n=1$ and subject to the stability of all other queues. The optimization problem is stated as follows:
\begin{equation}
\begin{split}
     &\underset{\substack{ {0\le\alpha_{\rm sr_1},\alpha_{\rm sr_2}\le1}\\{0\le a_{\rm s_1},a_{\rm s_2},a_{\rm sr_1},a_{\rm sr_2}\le 1}\\{0\le\eta_n\le1 \,\,\ \forall \ n\in\{1,2,3,4\}}}}{\max} \lambda_{\rm s}\!= \mu_{\rm s} \\
     &{\rm s.t.} \,\,\,\,\,\,\,\  a_{\rm s_1}+a_{\rm sr_1}=1,\ a_{\rm s_2}+a_{\rm sr_2}=1, \ \sum_{n=1}^4\eta_n=1 \\
     &\lambda_{\rm p_1}\!\le \! \mu_{\rm p_1},\  \lambda_{\rm p_2}\!\le\! \mu_{\rm p_2},\  \lambda_{\rm sr_1}\le \mu_{\rm sr_1},\ \lambda_{\rm sr_2}\le \mu_{\rm sr_2}.
     \label{optprob1}
     \end{split}
    \end{equation}
The optimization problem in~\eqref{optprob1} is nonconvex, however, it can be shown that, on fixing $(\alpha_{\rm sr_1},\alpha_{\rm sr_2})$, the optimization problem becomes a linear program, which is convex. Hence, it can be solved efficiently and reliably.

For a fixed $(\alpha_{\rm sr_1},\alpha_{\rm sr_2})$, $\pi_{\rm p_1}$ and $\pi_{\rm p_2}$ are constants. The objective function of \eqref{optprob1} after straightforward simplification becomes
\begin{equation}
\begin{split}
g&=\eta_1  \overline{P}_{\rm  s \rightarrow sd,W} \!+\!  \eta_2 \overline{P}_{\rm  s \rightarrow sd,W_{\rm p_1}} \!+\! \eta_3  \overline{P}_{\rm  s \rightarrow sd,W_{\rm p_2}} \\
& \ \!+\! \frac{\overline{\pi_{\rm p_2}}}{\pi_{\rm p_2}} a_{\rm s_1}  \overline{P}_{\rm  s \rightarrow sd,W_{\rm p_1}} \! +\!  \frac{\overline{\pi_{\rm p_1}}}{ \pi_{\rm p_1}} a_{\rm s_2} \overline{P}_{\rm  s \!\rightarrow \!sd,W_{\rm p_2}}.
\end{split}
\end{equation}

The stability of the ${\rm m}${\it th} primary queue provides a constraint on $\alpha_{\rm sr_m}$. Specifically,
\begin{equation}
\begin{split}
\alpha_{\rm sr_m}\ge \frac{\lambda_{\rm p}-\overline{P}_{\rm  p_m\rightarrow pd_m,W_{\rm p_m}}}{P_{\rm  p_m\rightarrow pd_m,W_{\rm p_m}} \overline{P}_{\rm  p_m \rightarrow s,W_{\rm p_m}}}.
\end{split}
\end{equation}
Adding the constraint that $\alpha_{\rm sr_m}\in[0,1]$, $\alpha_{\rm sr_m}$ is bounded by
\begin{equation}
\begin{split}
 \max\Big\{\frac{\lambda_{\rm p}-\overline{P}_{\rm  p_m\rightarrow pd_m,W_{\rm p_m}}}{P_{\rm  p_m\rightarrow pd_m,W_{\rm p_m}} \overline{P}_{\rm  p_m \rightarrow s,W_{\rm p_m}}},0\Big\}\le \alpha_{\rm sr_m}\le 1.
\end{split}
\end{equation}
 The constraints of the relaying queues are linear and can be simplified as follows:
\begin{equation}
\begin{split}
1-\frac{\lambda_{\rm sr_1}}{\pi_{\rm p_1} \overline{P}_{\rm  s \rightarrow pd_1,W_{\rm p_1}}}\ge \! \    \overline{\pi_{\rm p_2}} a_{\rm s_1}+\pi_{\rm p_2} \ (\eta_1+\eta_3),
\end{split}
\end{equation}

\begin{equation}
\begin{split}
1-\frac{\lambda_{\rm sr_2}}{\pi_{\rm p_2}\overline{P}_{\rm  s \rightarrow pd_2,W_{\rm p_2}}}&\ge\! \  \overline{\pi_{\rm p_1}} a_{\rm s_2}+\pi_{\rm p_1} \ (\eta_1+\eta_2),
\end{split}
\end{equation}
and

\begin{equation}
\begin{split}
\eta_1+\eta_2+\eta_3+\eta_4 \!=\! 1.
\end{split}
\end{equation}
Since the objective function and the constraints are linear, the optimization is a linear program, and thus can be solved efficiently and reliably \cite{boyed}.

For further simplification of the problem, we can assume that the secondary user accepts an equal fraction from each primary user. Specifically, $\alpha_{\rm sr_1}\!=\!\alpha_{\rm sr_2}\!=\!\alpha_{\rm sr}$. Accordingly, the problem reduces to a family of linear optimization problems parameterized by $\alpha_{\rm sr}$ which can be obtained by a simple grid search. The optimal $\alpha_{\rm sr}$ is taken as the one that yields the highest objective function.

If the primary users are symmetric, we have the following identities: $b_{\rm p_1}\!=\!b_{\rm p_2}\!=\!b$, $\lambda_{\rm p_1}\!=\!\lambda_{\rm p_2}\!=\!\lambda_{\rm p}$, $\alpha_{\rm sr_1}=\alpha_{\rm sr_2}=\alpha_{\rm sr}$, ${\rm W_{\rm p_1}=W_{\rm p_2}=W_{\rm p}\!=\!W/2}$, $\pi_{\rm p_1}\!=\!\pi_{\rm p_2}\!=\!\pi_{\rm p}$, $a_{\rm s_1}\!=\!a_{\rm s_2}\!=\!a_{\rm s}$ and $\eta_2=\eta_3=\eta$. The objective function becomes

\begin{equation}
\begin{split}
g&=\eta_1  \overline{P}_{\rm  s \rightarrow sd,W} \!+\!  \frac{2\overline{P}_{\rm  s \rightarrow sd,W/2}}{\pi_{\rm p}}\Bigg[\eta \pi_{\rm p} \ \!+\! \overline{\pi_{\rm p}} a_{\rm s}  \Bigg].
\end{split}
\end{equation}
The constraints become
\begin{equation}
\begin{split}
\eta_1+2\eta+\eta_4 \!=\! 1.
\end{split}
\end{equation}
and
\begin{equation}
\begin{split}
1-\frac{\lambda_{\rm sr}}{\pi_{\rm p} \overline{P}_{\rm  s \rightarrow pd,W/2}}\ge \! \    \overline{\pi_{\rm p}} a_{\rm s}\!+\!\eta\pi_{\rm p}\!+\! \eta_1\pi_{\rm p}.
\label{ffgg}
\end{split}
\end{equation}
Since we have a linear program, the optimal solution exists at the edge points of the feasible region. Let $y\!=\!\overline{\pi_{\rm p}} a_{\rm s}\!+\!\eta\pi_{\rm p}$ and $z=\pi_{\rm p} \eta_1$, where $y\in[0,1-\frac{\pi_{\rm p}}{2}]$ and $z\in[0,\pi_{\rm p}]$. We also define $\mathcal{D}\!=\frac{\lambda_{\rm sr}}{\pi_{\rm p} \overline{P}_{\rm  s \rightarrow pd,W/2}}$. The optimization problem in terms of $y$ and $z$ can be rewritten as follow:
\begin{equation}
\begin{split}
     &\underset{\substack{ {z\in[0,\pi_{\rm p}],y\in[0,1-\frac{\pi_{\rm p}}{2}]}\\{0\le a_{\rm s}\le 1}}}{\max.} z \!+\!  2\delta y \\ & \,\,\,\,\,\,\,\,\,\,\,\ {\rm s.t.} \,\,\,\,\,\,\,\   y\!+\! z\!\le\!1-\mathcal{D}, \\  & \,\,\,\,\,\,\,\,\,\,\,\,\,\,\,\,\,\,\,\,\,\,\,\,\,\,\,\,\  z+2y\le \pi_{\rm p}\!+\!2\overline{\pi_{\rm p}} a_{\rm s}-\eta_4,
     \end{split}
    \end{equation}
where $\delta\!=\!\frac{\overline{P}_{\rm  s \rightarrow sd,W/2}}{\overline{P}_{\rm  s \rightarrow sd,W}}$. Noting that $a_{\rm s}$ and $\eta_4$ appear only in the second constraint, setting $\!2\overline{\pi_{\rm p}} a_{\rm s}-\eta_4$ to its highest value, when $a_{\rm s}\!=\!1$ and $\eta_4=0$, will expand the feasible set (feasible region). The solution of the new optimization problem (\ref{ffggd}) is achievable by the original problem by setting $a_{\rm s}\!=\!1$ and $\eta_4=0$. Hence, solving the new optimization problem provides exactly the same solution of the original problem. The optimization problem after making the aforementioned change becomes
\begin{equation}
\begin{split}
     &\underset{\substack{ {z\in[0,\pi_{\rm p}],y\in[0,1-\frac{\pi_{\rm p}}{2}]}\\{0\le a_{\rm s}\le 1}}}{\max.} \mathcal{F}=z \!+\!  2\delta y \\ & \,\,\,\,\,\,\,\,\,\,\,\ {\rm s.t.} \,\,\,\,\,\,\,\   z\!+\! y\!\le\!1-\mathcal{D},\ z+2y\le 1\!+\!\overline{\pi_{\rm p}}.
     \label{ffggd}
     \end{split}
    \end{equation}
         \begin{figure}
    \centering
  \includegraphics[width=1\columnwidth]{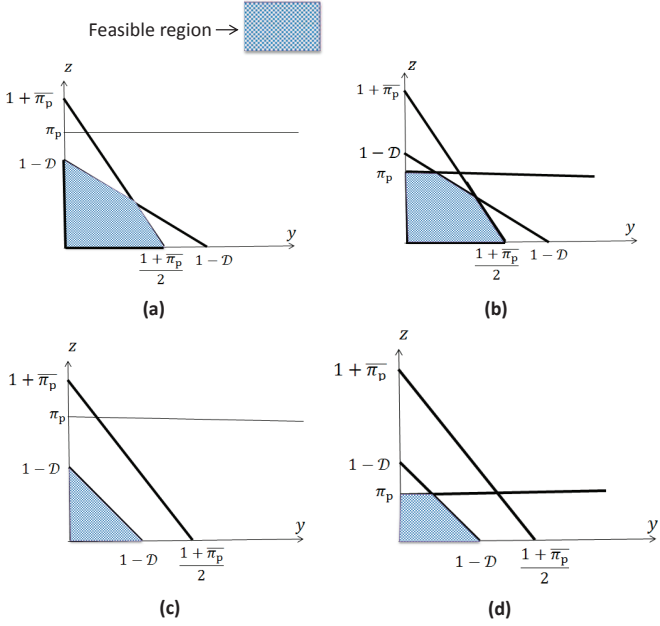}\\
  \caption{Feasible region for different scenarios. (a) feasible region for the case $1\!-\!\mathcal{D}\!\ge\!   \frac{1\!+\!\overline{\pi_{\rm p}}}{2}$ and $1-\mathcal{D}\le  \pi_{\rm p}$, (b) feasible region for the case $1-\mathcal{D}\ge   \frac{1\!+\!\overline{\pi_{\rm p}}}{2}$ and $1-\mathcal{D}>  \pi_{\rm p}$, (c) feasible region for the case $1-\mathcal{D}<   \frac{1\!+\!\overline{\pi_{\rm p}}}{2}$ and $1-\mathcal{D}< \pi_{\rm p}$, and (d) feasible region for the case $1-\mathcal{D}<  \frac{1\!+\!\overline{\pi_{\rm p}}}{2}$ and $1-\mathcal{D}> \pi_{\rm p}$.}\label{feas}
  \end{figure}
   Note that $\mathcal{D}$ must be strictly less than $1$ for the stability of the relaying queues to be attained and $1-\mathcal{D}\le   1\!+\!\overline{\pi_{\rm p}}$. The optimal solution of the optimization problem is obtained graphically. The feasible region for different conditions is shown in Fig. \ref{feas}. The main results are summarized as follows. If $1\!-\!\mathcal{D}\!\ge\!   \frac{1\!+\!\overline{\pi_{\rm p}}}{2}\rightarrow \frac{\mathcal{D}}{\pi_{\rm p}}\!\le\! 1/2$ and $1-\mathcal{D}\le  \pi_{\rm p}$, the optimal solution can be one of the following points (and perhaps the line segment connecting two points of them based on the value of $\delta$): $(y,z)=(0,1-\mathcal{D}), (\frac{1\!+\!\overline{\pi_{\rm p}}}{2},0)$ or the intersection point between the two lines $ z\!+\! y\!=\!1-\mathcal{D}$ and $\ z+2y= 1\!+\!\overline{\pi_{\rm p}}$ which is given by $(\overline{\pi_{\rm p}}+\mathcal{D},\pi_{\rm p}-2\mathcal{D})$. The optimal access probabilities for the first point are: $\eta_1=\frac{1-\mathcal{D}}{\pi_{\rm p}}$, $a^*_{\rm s}\!=\!0$, $\eta^*_4=0$, $\eta^*=0$. The optimal access probabilities for the second point are: $a^*_{\rm s}\!=\!1$, $a^*_{\rm sr}\!=\!0$, $\eta^*_1\!=\!0$, $\eta^*_4=0$, $\eta^*=1/2$. Finally, the optimal access probabilities for the third point are: $a^*_{\rm s}\!=\!1$, $a^*_{\rm sr}\!=\!0$, $\eta^*_1\!=\!1-2\frac{\mathcal{D}}{\pi_{\rm p}}$, $\eta^*_4=0$, $\eta^*=\frac{\mathcal{D}}{\pi_{\rm p}}$.

    If $1-\mathcal{D}\ge   \frac{1\!+\!\overline{\pi_{\rm p}}}{2}$ and $1-\mathcal{D}>  \pi_{\rm p}$, the optimal solution can be one of the following points (and perhaps the line segment connecting two points of them based on the value of $\delta$): $(y,z)=(0,\pi_{\rm p}), (1-\mathcal{D}-\pi_{\rm p},\pi_{\rm p}), (\frac{1\!+\!\overline{\pi_{\rm p}}}{2},0)$ or the intersection point $(\overline{\pi_{\rm p}}+\mathcal{D},\pi_{\rm p}-2\mathcal{D})$. For the first point $(0,\pi_{\rm p})$, the optimal access probabilities are: $a^*_{\rm s}\!=\!0$, $a^*_{\rm sr}\!=\!1\!$, $\eta^*_1\!=\!1$, $\eta^*_4=0$, $\eta^*\!=\!0$. For the second point $(1-\mathcal{D}-\pi_{\rm p},\pi_{\rm p})$, the optimal access probabilities are: $a^*_{\rm s}\!=\!1\!-\!\frac{\mathcal{D}}{\overline{\pi_{\rm p}}}$, $a^*_{\rm sr}\!=\!\frac{\mathcal{D}}{\overline{\pi_{\rm p}}}$, $\eta^*_1\!=\!1$, $\eta^*_4=0$, $\eta^*\!=\!0$. The optimal access probabilities for the third point $(\frac{1\!+\!\overline{\pi_{\rm p}}}{2},0)$ are: $a^*_{\rm s}\!=\!1$, $a^*_{\rm sr}\!=\!0$, $\eta^*_1\!=\!0$, $\eta^*_4=0$, $\eta^*=1/2$. Finally, the optimal access probabilities for the fourth point $(\overline{\pi_{\rm p}}+\mathcal{D},\pi_{\rm p}-2\mathcal{D})$ are: $a^*_{\rm s}\!=\!1$, $a^*_{\rm sr}\!=\!0$, $\eta^*_1\!=\!1-2\frac{\mathcal{D}}{\pi_{\rm p}}$, $\eta^*_4=0$, $\eta^*=\frac{\mathcal{D}}{\pi_{\rm p}}$. If  $1-\mathcal{D}<   \frac{1\!+\!\overline{\pi_{\rm p}}}{2}$ and $1-\mathcal{D}< \pi_{\rm p}$, the optimal solution can be one of the following points: $(y,z)=(0,1-\mathcal{D})$ or $(1-\mathcal{D},0)$. Under this case, for the first point $(0,1-\mathcal{D})$, the optimal access probabilities are: $a^*_{\rm s}\!=0$, $a^*_{\rm sr}\!=\!1$, $\eta_1\!=\!\frac{1-\mathcal{D}}{\pi_{\rm p}}$, $\eta^*_4=1-\frac{1-\mathcal{D}}{\pi_{\rm p}}$, $\eta^*\!=\!0$. For the second point $(1-\mathcal{D},0)$, the optimal access probabilities are:  $a^*_{\rm s}\!=\!\frac{1-\mathcal{D}-\pi_{\rm p}\eta}{\overline{\pi_{\rm p}}}$, $a^*_{\rm sr}\!=\!1\!-\!a^*_{\rm s}$, $\eta^*_1\!=\!0$, $\eta^*_4=1\!-\!2\eta^*$, $\eta^*\!=\![(\pi_{\rm p}-\mathcal{D})/\pi_{\rm p},1/2]$. Note that $(\pi_{\rm p}-\mathcal{D})/\pi_{\rm p}$ is always less than $1/2$ because $1-\mathcal{D}<  \frac{1\!+\!\overline{\pi_{\rm p}}}{2}$. If $1-\mathcal{D}<  \frac{1\!+\!\overline{\pi_{\rm p}}}{2}$ and $1-\mathcal{D}> \pi_{\rm p}$, the optimal solution can be one of the following points: $(y,z)=(0,\pi_{\rm p})$, $(1-\mathcal{D}-\pi_{\rm p},\pi_{\rm p})$ or $(1-\mathcal{D},0)$. For the first point $(0,\pi_{\rm p})$, the optimal access probabilities are: $a^*_{\rm s}\!=\!0$, $a^*_{\rm sr}\!=\!1\!$, $\eta^*_1\!=\!1$, $\eta^*_4=0$, $\eta^*\!=\!0$. For the second point $(1-\mathcal{D}-\pi_{\rm p},\pi_{\rm p})$, the optimal access probabilities are: $a^*_{\rm s}\!=\!1\!-\!\frac{\mathcal{D}}{\overline{\pi_{\rm p}}}$, $a^*_{\rm sr}\!=\!\frac{\mathcal{D}}{\overline{\pi_{\rm p}}}$, $\eta^*_1\!=\!1$, $\eta^*_4=0$, $\eta^*\!=\!0$. For the third point $(1-\mathcal{D},0)$, the optimal access probabilities are: $a^*_{\rm s}\!=\!\frac{1-\mathcal{D}-\pi_{\rm p}\eta}{\overline{\pi_{\rm p}}}$, $a^*_{\rm sr}\!=\!1\!-\!a^*_{\rm s}$, $\eta^*_1\!=\!0$, $\eta^*_4=1\!-\!2\eta^*$, $\eta^*\!=\![(\pi_{\rm p}-\mathcal{D})/\pi_{\rm p},1/2]$. Note that $(\pi_{\rm p}-\mathcal{D})/\pi_{\rm p}$ is always less than $1/2$ because $1-\mathcal{D}<  \frac{1\!+\!\overline{\pi_{\rm p}}}{2}$.

    The optimal solution of the problem is obtained as follows. For each $\alpha_{\rm sr}$, we compute $1-\mathcal{D}$, $\pi_{\rm p}$ and $\frac{1\!+\!\overline{\pi_{\rm p}}}{2}$. Based on the relationship between them, we get the candidate optimal points. Then, we substitute the optimal points into the objective function  $\mathcal{F}$. The point with the highest $ \mathcal{F}$ is taken as the optimal point which corresponds to $\alpha_{\rm sr}$. Afterwards, we compute $a^*_{\rm s}\!$, $a^*_{\rm sr}\!=\!1\!-\!a^*_{\rm s}$, $\eta^*_1$, $\eta^*_4$, $\eta^*$. Finally, we make a grid search over $\alpha_{\rm sr}$ in the original optimization problem (\ref{optprob1}) [under symmetry parameters] to find the optimal $\alpha_{\rm sr}$. The optimal $\alpha_{\rm sr}$ is taken as the one that yields the highest objective function in (\ref{optprob1}).

    The optimization problem (\ref{optprob1}) is solved at the cognitive radio terminal. It should be pointed out that the optimal parameters are functions of the mean arrival rates of the primary users and the channel outages which, in turn, as in (\ref{choutage}) are functions of many factors such as the received SNR at the receiving nodes, channel variance, transmission bandwidth, and packets size.

    \section{Numerical Results}\label{sec3}
\begin{figure}
    \centering
  \includegraphics[width=1\columnwidth]{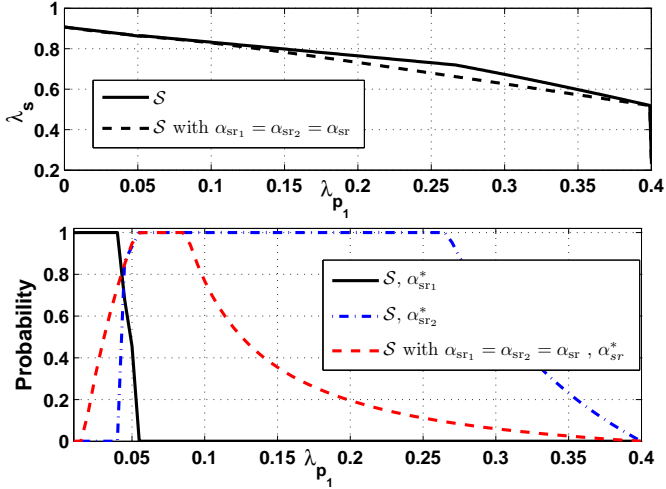}\\
  \caption{Maximum stable throughout and acceptance factor of the secondary user in the proposed cooperative protocol.}\label{fig1}
  \end{figure}

    In this section, we present some numerical results for the presented optimization problems in this paper. Let $\mathcal{S}$ denote the proposed cooperative system. For all figures, we choose $\tau$ to be a relatively large fraction of the time slot to validate the assumption of the perfect sensing. We first investigate the impact of setting $\alpha_{\rm sr_1}=\alpha_{\rm sr_2}\!=\!\alpha_{\rm sr}$ on the stability region. As seen in Fig. \ref{fig1}, the degradation of the secondary throughput due to using equal acceptance fractions is small. The figure also shows the non-increasing trend of the secondary throughput with $\lambda_{\rm p_1}$ for a fixed $\lambda_{\rm p_2}$. The parameters used to generate the figure are chosen to be: $\lambda_{\rm p_2}=0.2$ packets per time slot, $b_{\rm p_1}\!=\!b_{\rm p_2}=b_{\rm s}\!=\!1000$ bits, $\overline{P}_{\rm  p_1 \!\rightarrow \!pd_1,W_{\rm p_1}}\!=\!0.4$, $\overline{P}_{\rm  p_2 \!\rightarrow \!pd_2,W_{\rm p_2}}\!=\!0.5$, $\overline{P}_{\rm  s \!\rightarrow \!pd_1,W_{\rm p_1}}\!=\!0.1$, $\overline{P}_{\rm  s \!\rightarrow \!pd_2,W_{\rm p_2}}\!=\!0.3$, $\overline{P}_{\rm  p_1 \!\rightarrow \!s,W_{\rm p_1}}\!=\!\overline{P}_{\rm  p_2 \!\rightarrow \!s,W_{\rm p_2}}\!=\!0.6$, $\sigma^2_{\rm s,sd}\gamma_{\rm s,sd}=3.2$, ${\rm W}_{\rm p_1}=W_{\rm p_2}=2$ MHz, $T=1$ ms and $\tau=0.1T$.

    For comparison purposes, we introduce two systems that have a priority structure. The optimal acceptance fractions that achieve the boundary points of the stability region are also determined. Let $\mathcal{S}_1$ and $\mathcal{S}_2$ denote systems in which the secondary user cooperatively relays the undelivered primary packet while a priority in transmission is given to the relaying queues, i.e., $Q_{\rm s}$ is served when both relaying queues are emptied; and $\alpha_{\rm sr_1}=\alpha_{\rm sr_2}=1$, which means that the secondary terminal always accept the undelivered primary packets. The essential difference between $\mathcal{S}_1$ and $\mathcal{S}_2$ is that in system $\mathcal{S}_1$ the cognitive radio user employs our proposed band emerging technique with a strict priority given to serving relaying queues whenever possible, while system $\mathcal{S}_2$ can serve the secondary queue only when all primary and relaying queues are empty.

\begin{figure}
  \centering
  \includegraphics[width=1\columnwidth]{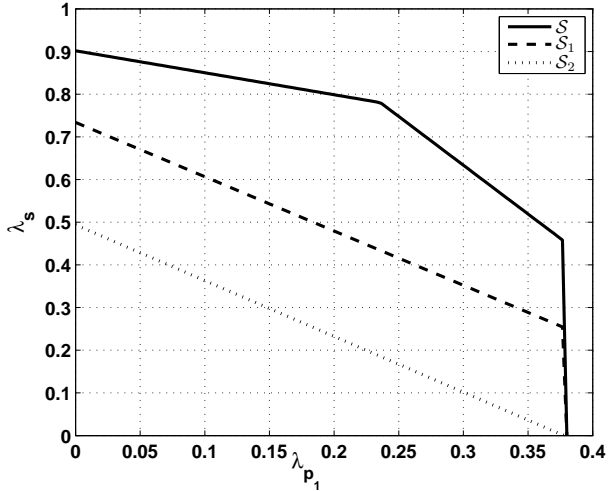}\\
  \caption{The maximum secondary stable throughput for $\mathcal{S}$, $\mathcal{S}_1$ and $\mathcal{S}_2$.}\label{fig3}
  \end{figure}

\begin{figure}
  \centering
  \includegraphics[width=1\columnwidth]{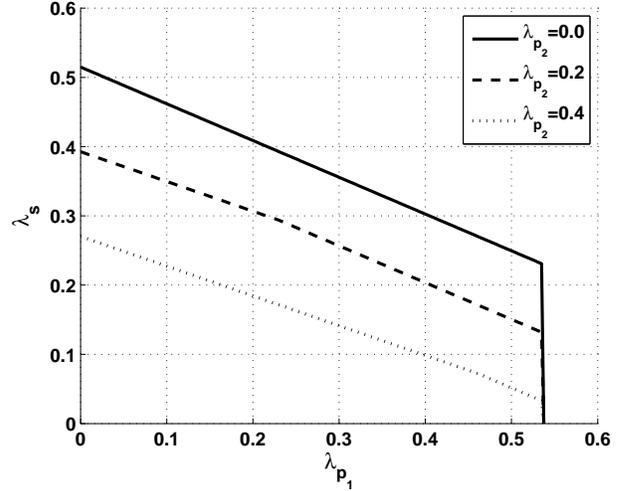}\\
  \caption{Stable throughput of the secondary user versus $\lambda_{\rm p_1}$ for different values of $\lambda_{\rm p_2}$.
  }\label{fig2}
  \end{figure}

    In Fig. \ref{fig3}, we show the maximum stable throughput of the secondary user for $\mathcal{S}_1$ and $\mathcal{S}_2$. From the figures, the maximum stable throughput for the cognitive radio user decreases with increasing arrival rates of the primary queues. The advantage of the proposed system over $\mathcal{S}_1$ and $\mathcal{S}_2$ is noted. Note that for the used parameters, when we set $\alpha_{\rm sr_1}=\alpha_{\rm sr_2}=\alpha_{\rm sr}$, we obtain exactly the same stability region of system $\mathcal{S}$. From the figure, we can infer that assigning the priority of transmission to the relaying queues causes secondary throughput degradation. The parameters used to generate the figure are: $\lambda_{\rm p_2}=0.2$ packets per time slot, $b_{\rm p_1}\!=\!b_{\rm p_2}=b_{\rm s}\!=\!1000$ bits, $\overline{P}_{\rm  p_1 \!\rightarrow \!pd_1,W_{\rm p_1}}\!=\!0.2$, $\overline{P}_{\rm  p_2 \!\rightarrow \!pd_2,W_{\rm p_2}}\!=\!0.3$, $\overline{P}_{\rm  s \!\rightarrow \!pd_1,W_{\rm p_1}}\!=\!0.6$, $\overline{P}_{\rm  s \!\rightarrow \!pd_2,W_{\rm p_2}}\!=\!0.6$, $\overline{P}_{\rm  p_1 \!\rightarrow \!s,W_{\rm p_1}}\!=\!\overline{P}_{\rm  p_2 \!\rightarrow \!s,W_{\rm p_2}}\!=\!0.6$, $\sigma^2_{\rm s,sd}\gamma_{\rm s,sd}=3.2$, ${\rm W}_{\rm p_1}=W_{\rm p_2}=2$ MHz, $T=1$ ms and $\tau=0.1T$.  Without cooperation, the maximum feasible arrival rate for ${\rm p_1}$ is $\overline{P}_{\rm  p_1 \!\rightarrow \!pd_1,W_{\rm p_1}}\!=\!0.2$ packets per slot. While, the maximum feasible arrival rate with cooperation is $0.38$ packets per time slot. Fig. \ref{fig2} demonstrates the stability region of the primary user ${\rm p_1}$ and the secondary user for different mean arrival rates of the primary user ${\rm p_2}$. We note that as the arrival rate of any of the primary queues increases, the maximum achievable secondary throughput decreases. The parameters used to generate the figure are given in Table \ref{table1}.

\section{Conclusion}\label{sec4}

   In this paper, we have investigated the stability region of a cognitive radio scenario incorporating two primary users and a cooperative cognitive user. We have proposed a novel medium access control protocol in which a single cognitive radio user can efficiently relay packets for two primary nodes with a dedicated relaying buffer at the secondary user for each primary user. The cognitive user may combine/merge the available primary orthogonal bands to increase the probability of successful packet reception, which in turn increases its service rate. Furthermore, the secondary user schedules its queues' access to the bands randomly based on the states of the primary queues. The access probability assigned to each queue depends on the joint state of the primary queues. The gains of our cooperative protocol is demonstrated relative to cognitive prioritized relaying scenarios, where the relaying queues have priority in transmission over the secondary queue. We have been able to model the throughput (stability) maximization problem as an optimization problem which has been formally solved. Our proposed protocol and analysis can be extended to a system composed of multiple cognitive radio users and primary users with sensing errors and concurrent transmission occurrence, which will be included in an extended future version of the this publication.

\begin{table}

\begin{center}
\caption{Parameters used to generate Fig. \ref{fig2}.}
\label{table1}
\begin{small}
\begin{tabular}{ |@{}c@{}|@{}c@{} |@{}c@{} |@{}c@{}|@{}c@{}|  }
    \hline
    $b_{\rm i}=1000 \,\,\ \forall {\rm i}$& ${\rm W}_{\rm p_2}=2$ MHz& ${\rm W}_{\rm p_1}=2$ MHz \\[3pt]\hline $\gamma_{\rm p_2,s}=7.5$ & $T=10^{-3}$ second & $\tau=0.1T$ \\[3pt]\hline $\sigma^2_{\rm p_1,pd_1}=0.04$ & $\sigma^2_{\rm p_2,pd_2}=0.02$& $\sigma^2_{\rm s,pd_1}=0.8$\\[3pt]\hline $\sigma^2_{\rm s,pd_2}=0.8$& $\sigma^2_{\rm s,sd}=0.8$& $\sigma^2_{\rm p_1,s}=0.7$\\[3pt]\hline $\sigma^2_{\rm p_2,s}=0.75$& $\gamma_{\rm p_1,pd_1}=8$& $\gamma_{\rm p_2,pd_2}=6$ \\[3pt]\hline $\gamma_{\rm s,pd_2}=6$& $\gamma_{\rm s,sd}=0.4$ & $\gamma_{\rm p_1,s}=7$\\[3pt]\hline ${\rm W=W_{\rm p_1}+W_{\rm p_2}}$& $\gamma_{\rm s,pd_1}=5$ & \\[5pt]\hline
\end{tabular}
\end{small}
\end{center}
\end{table}

\balance
\bibliographystyle{IEEEtran}
\bibliography{IEEEabrv,term_bib}

\begin{thebibliography}{1}
\providecommand{\url}[1]{#1}
\csname url@samestyle\endcsname
\providecommand{\newblock}{\relax}
\providecommand{\bibinfo}[2]{#2}
\providecommand{\BIBentrySTDinterwordspacing}{\spaceskip=0pt\relax}
\providecommand{\BIBentryALTinterwordstretchfactor}{4}
\providecommand{\BIBentryALTinterwordspacing}{\spaceskip=\fontdimen2\font plus
\BIBentryALTinterwordstretchfactor\fontdimen3\font minus
  \fontdimen4\font\relax}
\providecommand{\BIBforeignlanguage}[2]{{%
\expandafter\ifx\csname l@#1\endcsname\relax
\typeout{** WARNING: IEEEtran.bst: No hyphenation pattern has been}%
\typeout{** loaded for the language `#1'. Using the pattern for}%
\typeout{** the default language instead.}%
\else
\language=\csname l@#1\endcsname
\fi
#2}}
\providecommand{\BIBdecl}{\relax}
\BIBdecl

\bibitem{simeone}
O.~Simeone, Y.~Bar-Ness, and U.~Spagnolini, ``Stable throughput of cognitive
  radios with and without relaying capability,'' \emph{IEEE Trans. Commun.},
  vol.~55, no.~12, pp. 2351--2360, Dec. 2007.

\bibitem{khattab}
M.~Elsaadany, M.~Abdallah, T.~Khattab, M.~Khairy, and M.~Hasna, ``Cognitive
  relaying in wireless sensor networks: Performance analysis and
  optimization,'' in \emph{Proc. IEEE GLOBECOM}, Dec. 2010, pp. 1--6.

\bibitem{krikidis2009protocol}
I.~Krikidis, J.~Laneman, J.~Thompson, and S.~McLaughlin, ``Protocol design and
  throughput analysis for multi-user cognitive cooperative systems,''
  \emph{IEEE Trans. Wirel. Commun.}, vol.~8, no.~9, pp. 4740--4751, Sept. 2009.

\bibitem{close}
J.~Gambini, O.~Simeone, and U.~Spagnolini, ``Cognitive relaying and
  opportunistic spectrum sensing in unlicensed multiple access channels,'' in
  \emph{Proc. IEEE 10th ISSSTA}, Aug. 2008, pp. 371--375.

\bibitem{krikidis2010stability}
I.~Krikidis, N.~Devroye, and J.~Thompson, ``Stability analysis for cognitive
  radio with multi-access primary transmission,'' \emph{IEEE Trans. Wirel.
  Commun.}, vol.~9, no.~1, pp. 72--77, Jan. 2010.

\bibitem{bao2010stable}
X.~Bao, P.~Martins, T.~Song, and L.~Shen, ``Stable throughput analysis of
  multi-user cognitive cooperative systems,'' in \emph{Proc. IEEE GLOBECOM},
  Dec. 2010, pp. 1--5.

\bibitem{Sult1212:Cooperative}
A.~{El Shafie} and A.~Sultan, ``Cooperative cognitive relaying with ordered
  cognitive multiple access,'' in \emph{Proc. IEEE GLOBECOM}, Dec. 2012, pp.
  1434--1439.

\bibitem{sadek}
A.~Sadek, K.~Liu, and A.~Ephremides, ``Cognitive multiple access via
  cooperation: protocol design and performance analysis,'' \emph{IEEE Trans.
  Inf. Theory}, vol.~53, no.~10, pp. 3677--3696, Oct. 2007.

\bibitem{boyed}
S.~Boyd and L.~Vandenberghe, \emph{Convex {Optimization}}.\hskip 1em plus 0.5em
  minus 0.4em\relax Cambridge University Press, 2004.

\end{thebibliography}
\end{document}